# Reconfigurable Nonlocal Light-Emitting Metalens Nanolasers via Bound States in the Continuum

Omar A. M. Abdelraouf [1,*]

**Affiliations:**
[1]Institute of Materials Research and Engineering, Agency for Science, Technology, and Research (A*STAR), 2 Fusionopolis Way, #08-03, Innovis, Singapore 138634, Singapore

*Corresponding author: Omar A.M. Abdelraouf (email: Omar_Abdelrahman@a-star.edu.sg)

## ABSTRACT

The development of on-chip, reconfigurable coherent light sources operating in the visible spectrum is critical for next-generation optical computing and high-speed data processing. Here, we demonstrate an all-optically tunable nanolaser based on a hybrid dielectric metasurface. The platform integrates a Rhodamine B (Rd B) dye gain medium with a Niobium Pentoxide ($Nb_2O_5$) metasurface supporting a high-*Q* Bound State in the Continuum (BIC) resonance. Active, non-volatile tuning is achieved by incorporating the low-loss phase change material (PCM) Antimony Trisulfide ($Sb_2S_3$). We demonstrate room-temperature lasing at 616 nm with Q-factor > 2050. A continuous-wave (CW) laser is utilized for the crystallization of the PCM, tuning the lasing emission to crystalline at 621 nm with Q-factor > 1454, while a near-infrared (NIR) ps-laser triggers its re-amorphization. This robust, thermally protected optical-only control scheme, combined with a phase-gradient metasurface design, realizes a coherent, nonlocal light-emitting metalens for simultaneous lasing wavelength tuning and wavefront control. This work provides a foundational demonstration of an all-optical tunable metalens-based nanolaser for applications in high-speed optical communication, neuromorphic photonics, dynamic holographic displays, and reconfigurable quantum light sources.

## INTRODUCTION

The exponential growth in data processing and communication has exposed the fundamental limitations of electronic integrated circuits (EICs), which are increasingly constrained by resistive losses, electromagnetic interference, and signal delays.[1-3] Photonic integrated circuits (PICs) offer a disruptive alternative, leveraging photons as information carriers to achieve unprecedented data speeds, lower energy consumption, and massive bandwidth.[4-6] A central challenge in realizing large-scale, complex PICs is the development and dense integration of efficient, compact, on-chip coherent light sources.[7] Nanolasers, with their small footprints, low power consumption, and potential for high modulation speeds, are considered essential building blocks for this new generation of photonic processors.[8] However, integrating nanolasers remains a significant hurdle. Many nanolaser designs suffer from poor spatial coherence, acting as point sources that require complex off-chip optics to collect and collimate their emission, which undermines the primary goal of dense, on-chip integration.[9]

All-dielectric metasurfaces, planar arrays of subwavelength nanostructures, have emerged as a powerful platform to overcome these challenges by offering unprecedented control over light-matter interactions.[10-16] By avoiding the ohmic losses inherent to their plasmonic counterparts,[17-26] dielectric metasurfaces can engineer strong multipolar resonances to confine light at the nanoscale.[27-32] A particularly promising approach involves leveraging Bound States in the Continuum (BICs), which are resonant modes with theoretically infinite lifetimes that manifest in periodic structures as ultra-high-Q factor resonances.[33-35] This mechanism enables extreme local field enhancement and provides the strong optical feedback necessary for achieving low-threshold nanolasing from integrated gain materials.[36] Concurrently, metasurfaces provide a second, revolutionary capability: arbitrary wavefront manipulation. By imparting a spatially-varying phase shift to the incident light, phase-gradient metasurfaces can replace bulky, refractive optical components, enabling flat-optic devices such as metalenses, beam steerers, and holograms.[37-39] This dual ability to both generate light via enhanced light-matter interaction and control its propagation via wavefront shaping makes metasurfaces a uniquely suitable platform for advanced, integrated light-emitting devices.

Despite this promise, the functionality of most metasurface-based devices is static, determined irreversibly by the nanostructure's geometry during fabrication. This post-fabrication rigidity presents a significant bottleneck for practical photonic systems, which increasingly demand dynamic, tunable, and reconfigurable responses to handle complex, real-time tasks.[40] A truly "programmable" photonic chip requires components whose optical properties such as resonant wavelength, focal length, or beam direction can be actively modulated in situ. Reconfigurable metasurfaces, which integrate active materials to achieve dynamic control, have thus become a major research frontier. Various tuning mechanisms have been explored, including mechanical deformation,[41] thermal modulation,[42-44] and electrical gating using liquid crystals.[45] While effective, these methods often suffer from slow response times, volatile control (requiring constant power to maintain a state), or complex integration. Chalcogenide Phase Change Materials (PCMs) have emerged as a transformative solution, offering a unique combination of ultrafast (sub-nanosecond) switching speeds and, critically, non-volatile state retention.[46] This non-volatility, wherein the material holds its optical state even after the tuning stimulus is removed, is highly desirable for minimizing static power consumption. However, the most common PCMs, such as $Ge_2Sb_2Te_5$ (GST), are prohibitively lossy in the visible spectrum, limiting their use to infrared applications. This has spurred research into alternative low-loss PCMs, with antimony-based compounds like $Sb_2S_3$ and Antimony Triselenide ($Sb_2Se_3$) demonstrating a large refractive index contrast with sufficient transparency for reconfigurable visible-light nanophotonics.[16]

Herein, we bridge these fields by designing, fabricating, and experimentally demonstrating an all-optically tunable, visible-spectrum nanolaser. The device is a hybrid platform integrating an organic dye (Rhodamine B) gain medium with a high-Q dielectric metasurface cavity, which is patterned from Niobium Pentoxide ($Nb_2O_5$) to support a strong BIC resonance. Ultrafast, non-volatile active control is achieved by incorporating a low-loss $Sb_2S_3$ PCM layer. We demonstrate room-temperature lasing at 616 nm, which is all-optically tuned to 621 nm following a cw-laser opto-thermal crystallization. Notably, the re-amorphous phase tuning utilize ps-laser pulse (1060 nm) and decoupled from the gain medium's absorption (pumped at 532 nm), and a novel device architecture incorporating an $Al_2O_3$ thermal insulator and back-side illumination prevents thermal damage to the dye. Furthermore, by patterning the BIC metasurface with a spatial phase gradient metalens, we propose and demonstrate a coherent, nonlocal tunable light-emitting metalens, that will simultaneously generates coherent visible light and shapes its wavefront for wavelength tuning and focusing simultaneously. This work provides a fundamental building block for future ultrafast, reconfigurable nanophotonic systems, with direct applications in optical communication,[47] high-speed optical computing,[48] and active wavefront control.[49]

## RESULTS AND DISCUSSION

The realization of an on-chip, dynamically reconfigurable coherent light source requires a delicate balance between high-*Q* factor mode confinement, efficient gain integration, and a non-destructive active tuning mechanism. The proposed architecture, depicted in Figure 1a, addresses these concurrent requirements through a highly engineered hybrid metasurface platform operating in the visible spectrum. The fundamental cavity relies on $Nb_2O_5$, a low-loss ($k$~0.001) high-index dielectric ($n$>2.3) chosen for its exceptionally optical properties in the visible regime (Figure S1), avoiding the severe non-radiative ohmic damping that conventionally plagues plasmonic nanolasers. To achieve the requisite temporal confinement of photons for stimulated emission, the $Nb_2O_5$ layer is structured into a periodic array of asymmetric meta-atoms designed to support BIC resonance. Translating such intricate geometries from theoretical design to physical reality demands scalable, high-fidelity manufacturing protocols. The fabrication of these nanostructures relies on E-beam lithography combined with atomic layer deposition (ALD) and blanket reactive-ion etching (RIE) (See Methods section and Figure S2). This wafer-scale fabrication approach ensures the precise replication of the critical dimensions such as widths ($w_1$ and $w_2$), the gap ($D$) and maintaining perfectly vertical sidewalls, which is an absolute necessity to prevent unwanted out-of-plane scattering that would otherwise degrade the cavity *Q*-factor.

The integration of the active tuning medium and the gain material necessitates a multilayered approach to isolate their competing physical requirements. $Sb_2S_3$ layer is sputtered beneath the $Nb_2O_5$ resonators. $Sb_2S_3$ was strategically selected due to its wide bandgap, which provides sufficient transparency in the visible spectrum while exhibiting a massive refractive index contrast ($\Delta n$) between its amorphous and crystalline states. This index contrast provides the localized optical perturbation required to actively shift the BIC resonance. However, the opto-thermal switching of PCMs typically generates localized heat that can permanently photobleach or thermally degrade sensitive organic gain media like Rhodamine B (Rd B). To circumvent this, a 10 nm thick Aluminium Oxide ($Al_2O_3$) layer is deposited via ALD as a highly conformal thermal barrier. This dielectric spacer provides encapsulation protection and critical thermal boundary resistance, blocking the high-temperature PCM phase transition from the overlying Rd B dye layer while remaining optically thin enough to allow the evanescent tail of the BIC mode to overlap strongly with the gain medium.

The physical mechanism enabling lasing in this subwavelength architecture is governed by symmetry-protected BICs. A true BIC exists strictly at the Γ-point of the momentum space, where it is perfectly uncoupled from the radiative continuum, possessing a theoretically infinite $Q$-factor but zero coupling to external excitation. To harness this mode for lasing, the structural symmetry of the meta-atom must be intentionally broken. As illustrated in the simulated transmission spectra (Figure 1c), a highly symmetric unit cell ($w_1 = w_2$) exhibits flat transmission with no sharp resonant features, as the mode remains entirely dark. By introducing an asymmetry parameter, defined broadly as $\boldsymbol{\alpha} \propto \Delta w$, the true BIC collapses into a quasi-BIC. This topological transition allows the mode to couple weakly to the free-space radiative continuum, manifesting as an ultra-narrow Fano resonance near 616 nm when the $Sb_2S_3$ layer is in its amorphous state (a-$Sb_2S_3$). The radiative $Q$-factor of this quasi-BIC scales according to an inverse square law, $Q \propto \boldsymbol{\alpha}^{-2}$, allowing deterministic control over the cavity losses. The experimental validation of this phenomenon (Figure 1d) confirms the successful manifestation of the quasi-BIC, yielding a measured cold-cavity $Q$-factor of 242. While lower than the theoretical limit due to due to intrinsic material surface roughness, and finite array size effects this value represents a remarkably high optical feedback parameter for a visible-spectrum flat cavity, establishing a rigorous foundation for low-threshold lasing.

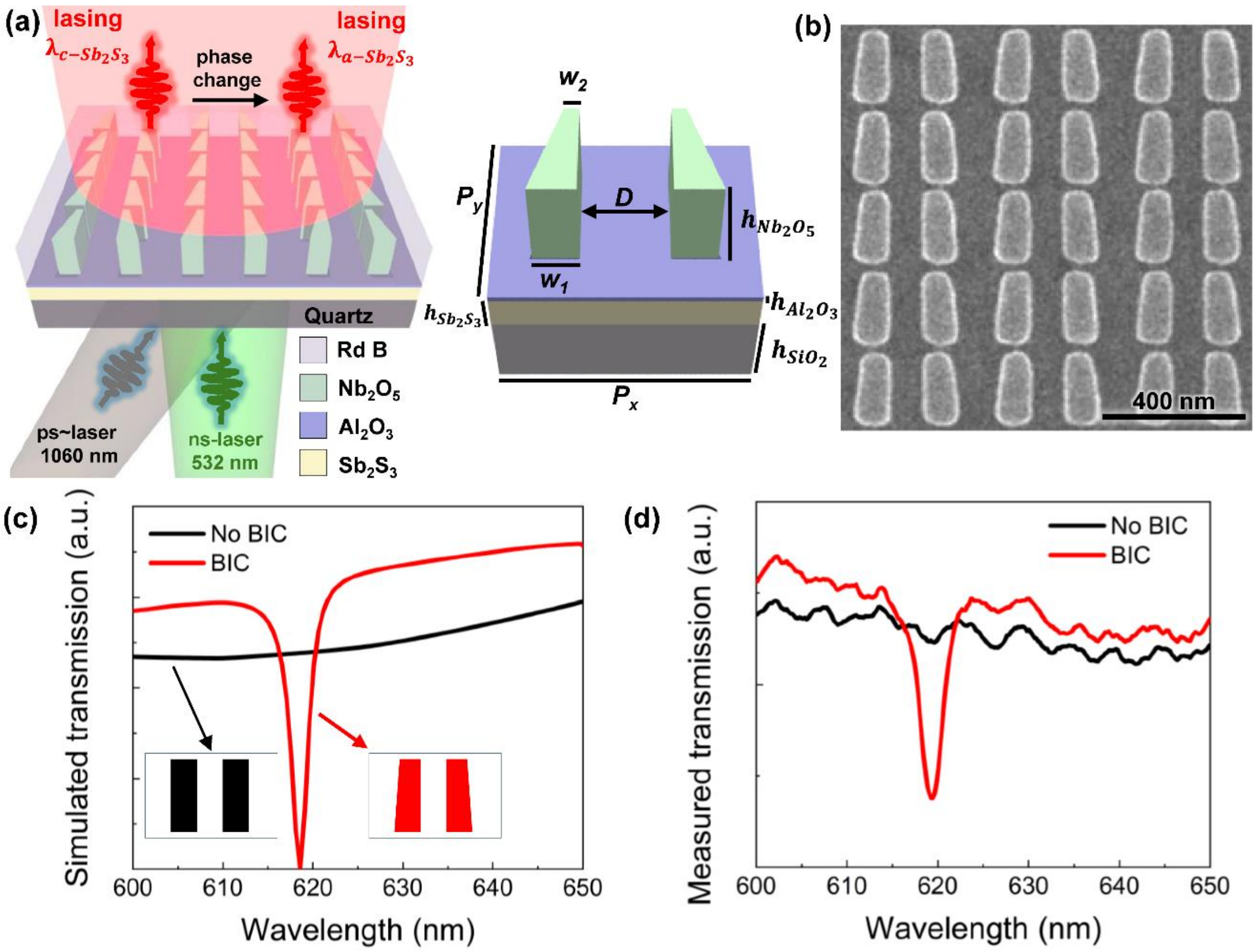

**Figure 1.** Design and working principle of the all-optically reconfigurable hybrid metasurface nanolaser operating via BIC resonances in the visible spectrum. (a) Three-dimensional representation of the device architecture. The platform combines a low-loss $Nb_2O_5$ BIC cavity, an $Al_2O_3$ thermal barrier, an $Sb_2S_3$ active tuning layer, and a Rhodamine B (Rd B) gain medium resting on a quartz substrate. It illustrates excitation via a 532 nm green nanosecond (ns) pump laser alongside the resulting tunable emission. (Inset) Perspective view of the hybrid unit cell detailing structural dimensions. Fixed dimensions include periods $P_x$ = 405 nm and $P_y$ = 234 nm; layer thicknesses for $Nb_2O_5$, $Sb_2S_3$, and $Al_2O_3$ are 450 nm, 130 nm, and 10 nm, respectively, on a 500 µm quartz base. Tunable geometric parameters comprise the widths ($w_1$, $w_2$) and the horizontal gap

(*D*). (b) Scanning electron micrograph (SEM) of the bare hybrid metasurface (prior to Rd B spin-coating), featuring a 200 nm scale bar. (c) Simulated transmission spectra illustrating the absence of a BIC mode in a symmetric design, compared to the emergence of a high-*Q* BIC resonance near 617 nm upon introducing structural asymmetry in the amorphous $Sb_2S_3$ (a-$Sb_2S_3$) state. (d) Experimental transmission spectra for the a-$Sb_2S_3$ phase, confirming the presence of a BIC resonance with a measured *Q*-factor of 242 when symmetry is broken.

To unravel the complex light-matter interactions and radiation suppression mechanisms driving the high-*Q* quasi-BIC, a comprehensive electrodynamic multipolar decomposition was performed. The normalized scattering cross-section (NSCS) spectra, calculated via the integration of polarization currents within the unit cell, decompose the far-field radiation into its fundamental Cartesian multipole moments: the electric dipole (*ED*), magnetic dipole (*MD*), electric quadrupole (*EQ*), and magnetic quadrupole (*MQ*). As shown in Figure 2a for the a-$Sb_2S_3$ state, the resonance is not dominated by a single localized mode but rather emerges from the destructive interference of multiple co-located radiative pathways. Specifically, the sharp quasi-BIC resonance is characterized by an abrupt suppression of the electric dipole radiation and a simultaneous coupling between magnetic dipole and electric quadrupole moments. The superposition of these multipoles creates a quasi-BIC state in the far-field, effectively trapping photons within the dielectric volume. When the underlying $Sb_2S_3$ layer is transitioned to the crystalline state (c-$Sb_2S_3$), the increased refractive index dramatically alters the local phase accumulation and effective index of substrate (Figure 2b). This opto-thermal perturbation modifies the coupling coefficients between the fundamental multipoles, red-shifting the spectral position of the destructive interference condition and dragging the quasi-BIC resonance to a longer wavelength, thereby proving the viability of the index-shifting mechanism on the multipolar balance of the cavity.

The spatial distribution of the electromagnetic fields provides further insight into the gain-cavity coupling efficiency. The simulated near-field electric intensity profiles at the BIC resonance (Figure 2c and 2d) reveal extreme light confinement. The $|E_{x\text{-}y}|^2$ and $|E_{x\text{-}z}|^2$ cross-sections demonstrate that the optical energy is heavily localized within the high-index $Nb_2O_5$ pillars, forming standing-wave patterns characterized by intense "hotspots" near the dielectric boundaries. Crucially, a significant fraction of this enhanced near-field extends as an evanescent wave into the adjacent Rd B gain layer. The efficiency of a nanolaser is fundamentally dictated by how effectively the stimulated emission is coupled into the cavity mode, a phenomenon heavily dependent on the spatial overlap and the Purcell effect. It is a critical distinction in nanophotonic cavity electrodynamics that the PL enhancement is not simply derived from the absolute minimization of the effective modal volume ($V_m$) alone. Rather, the overall PL enhancement is the resultant of the $Q$-factor enhancement over the modal volume, scaling strictly proportional to the ratio $Q/V_m$. By operating in the quasi-BIC regime, the metasurface simultaneously maximizes $Q/V_m$ ratio drastically elevates the spontaneous emission rate into the lasing mode, ensuring that a vast majority of the excited Rd B molecules contribute to the coherent oscillation rather than radiating isotropically into free space.

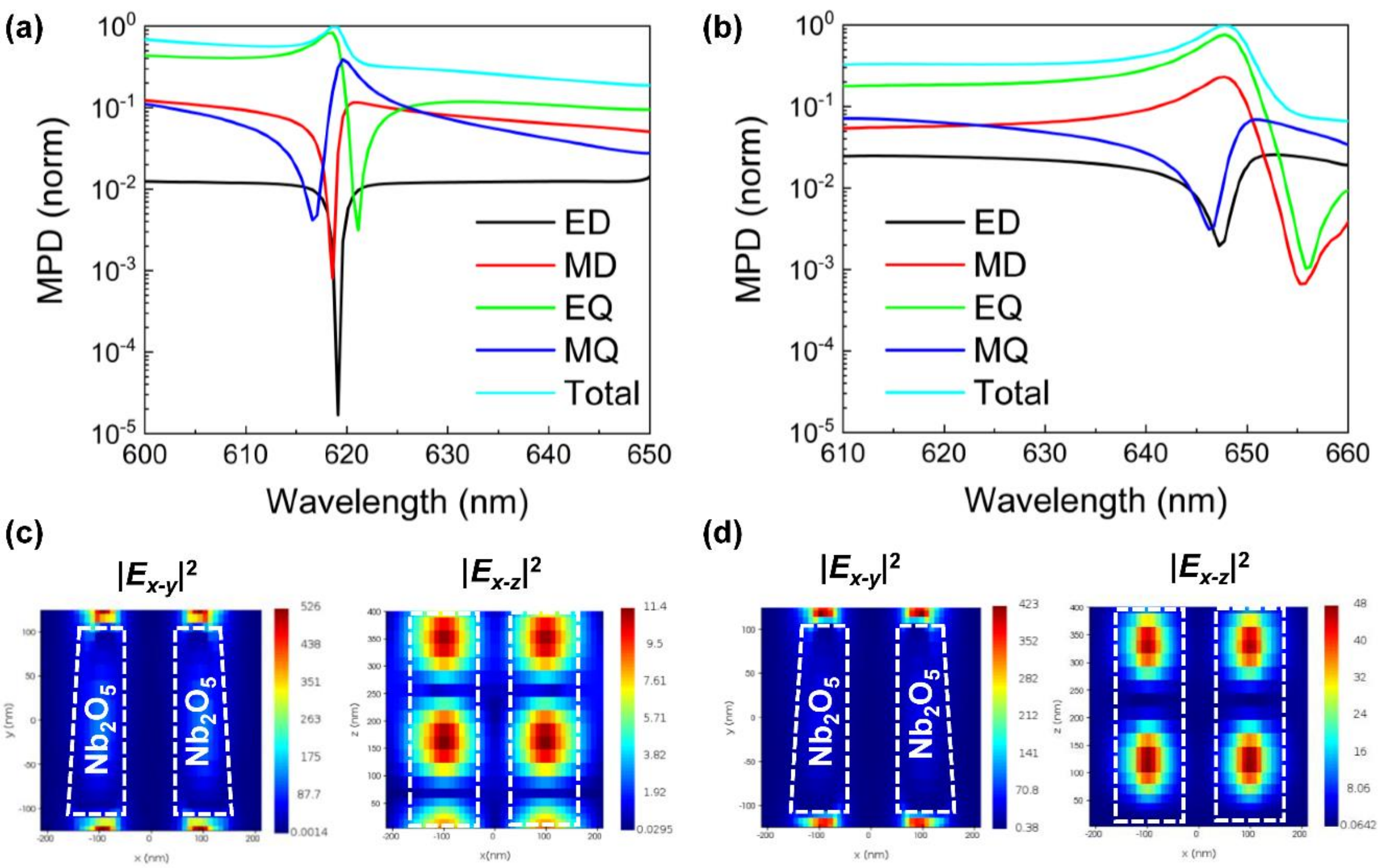

**Figure 2.** Electromagnetic field distributions and multipolar analysis of the reconfigurable hybrid nanolaser cavity. (a) Normalized scattering cross-section (NSCS) spectra for the asymmetric BIC metasurface in the amorphous $Sb_2S_3$ (a-$Sb_2S_3$) phase. The plot resolves the constituent electric, magnetic, and toroidal contributions, including electric dipole (*ED*), electric quadrupole (*EQ*), magnetic dipole (*MD*), and magnetic quadrupole (*MQ*) modes. (b) Wavelength-dependent NSCS profile of the hybrid BIC structure upon transition to the crystalline state (c-$Sb_2S_3$). (c) Simulated electric field intensity profiles ($|E_{x\text{-}y}|^2$) mapping the horizontal cross-section at the mid-height of the $Nb_2O_5$ resonators, alongside the vertical cross-section ($|E_{x\text{-}z}|^2$) spanning the hybrid pillars during BIC resonance excitation. (d) Corresponding electric field intensity mappings across both the horizontal ($|E_{x\text{-}y}|^2$) and vertical ($|E_{x\text{-}z}|^2$) planes under resonant excitation conditions, further illustrating the spatial confinement within the metasurface architecture.

The experimental realization of coherent emission was first characterized with the $Sb_2S_3$ tuning layer initialized in the as-deposited amorphous state. The hybrid metasurface was optically pumped using a low power 532 nm nanosecond (ns) pulsed laser, chosen to efficiently excite the absorption band of the Rd B dye while circumventing any two-photon absorption or thermal perturbation of the underlying PCM layer. Figure 3a illustrates the spectral evolution of the normalized PL as a function of the incident pump fluence. At low excitation fluences, the emission is characterized by a broad, low-intensity peak typical of spontaneous fluorescence. The spectral shape in this regime is largely governed by the intrinsic emission cross-section of the Rd B molecules, weakly modulated by the cold-cavity resonances. As the pump fluence is progressively increased, the carrier inversion density within the dye molecules surpasses the total cavity losses (comprising both intrinsic non-radiative decay and radiative leakage). Consequently, a highly localized, ultra-narrow emission peak rapidly emerges precisely at the wavelength of the quasi-BIC resonance at wavelength 616 nm, rapidly outcompeting the background spontaneous emission.

The transition dynamics from thermal light to coherent laser emission are definitively captured by analysing the light-in versus light-out (L-L) curve on a double-logarithmic scale, as presented in Figure 3b. The integrated PL intensity exhibits the classical non-linear transition response indicative of microcavity lasing. The lower linear region corresponds to the spontaneous emission regime, followed by a sharp super-linear kink identifying the lasing threshold of 175 $\mu J/cm^2$, where stimulated emission begins to dominate the photon generation process. Above this critical threshold fluence, the output intensity returns to a linear slope, confirming that the system is fully clamped in the lasing regime. Concurrently, the spectral full-width-half-maximum (FWHM) of the emission peak undergoes a dramatic collapse to 0.3 nm. Below threshold, the broad linewidth reflects the inhomogeneous broadening of the dye. At the onset of lasing, phase coherence is established, and the linewidth dramatically narrows, limited only by the cavity's passive bandwidth and temporal coherence fluctuations. From the narrowest experimentally resolved linewidth above threshold, an active lasing $Q$-factor exceeding 2050 was calculated ($Q = \lambda/\Delta\lambda$). This active $Q$-factor represents a significant enhancement over the cold-cavity measurement, primarily because the introduction of optical gain actively compensates for the absorptive losses of the materials. The exceptionally high $Q$-factor measured at room temperature validates the quasi-BIC architecture as a highly efficient feedback mechanism, capable of sustaining robust coherence in lossy visible-spectrum organic gain media without the onset of severe fluorescence quenching.

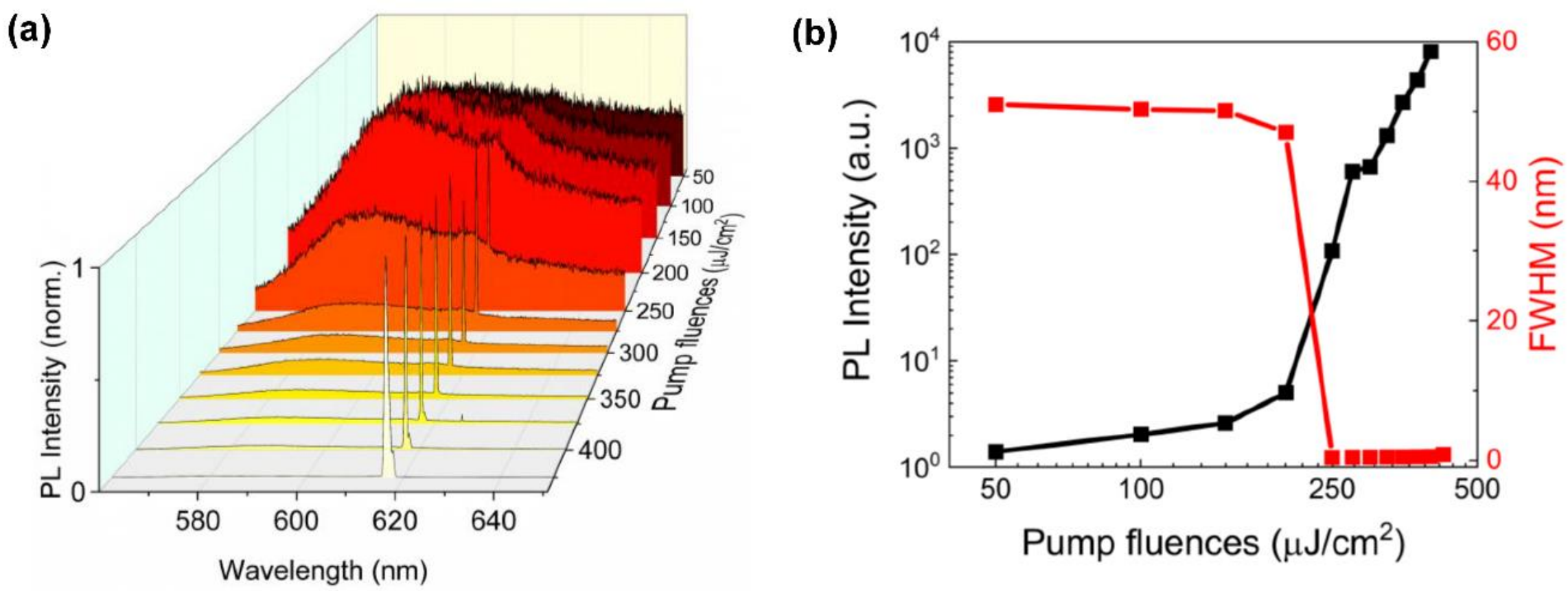

**Figure 3.** Emission characteristics of the tunable hybrid BIC nanolaser in the amorphous phase (a-$Sb_2S_3$). (a) Experimental normalized photoluminescence (PL) spectra as a function of pump fluence for the Rd B-coated hybrid metasurface. The data captures the distinct transition from broad spontaneous emission to narrow lasing action precisely at the cavity's BIC wavelength. (b) Integrated PL intensity and spectral full-width-half-maximum (FWHM) plotted against pump fluence for the 616 nm emission peak. The clear change in slope on the logarithmic scale, coupled with a sharp drop in FWHM, confirms the onset of lasing action with a calculated $Q$-factor exceeding 2050.

The most significant advancement presented in this architecture is the demonstration of non-volatile and all-optical tuning of the lasing emission, fundamentally moving beyond static metasurface designs. The spectral tuning is achieved by harnessing the drastic, non-volatile structural phase transition of the embedded $Sb_2S_3$ layer. As shown in the main panel of Figure 4, continuous ns-laser pumping above the lasing threshold yields a stable coherent peak near 616 nm when the PCM is in the a-$Sb_2S_3$ state. To trigger the spectral shift, a carefully engineered two-pulse optical control scheme is employed. The crystallization process is induced via localized opto-thermal heating using a focused CW-laser (405 nm). This wavelength is highly absorbed by the $Sb_2S_3$ layer, slowly elevating the local temperature above the glass transition point ($T_g$) but keeping it safely below the melting temperature ($T_m$). As the atomic lattice rearranges into the more ordered c-$Sb_2S_3$ state, the real part of the refractive index experiences a substantial increase. Because the quasi-BIC mode profile exhibits significant spatial overlap with the PCM layer, this localized index modulation alters the effective optical index of substrate and the phase matching conditions of the cavity. Consequently, the lasing emission is dynamically tuned, red-shifting to 621 nm. Notably, the active $Q$-factor in the crystalline state remains high (>1454), proving that the $Sb_2S_3$ material maintains sufficiently low optical losses in the visible spectrum even after crystallization, a feat that is entirely impossible with traditional PCMs like $Ge_2Sb_2Te_5$ (GST) which become highly metallic and absorptive in the visible regime. The high spatial precision of this all-optical tuning is corroborated by the WiTec optical microscopy image (Figure 4, inset). The well-defined boundary between the highly reflective switched c-$Sb_2S_3$ area and the unswitched a-$Sb_2S_3$ background demonstrates that the thermal diffusion is tightly confined. This localized switching implies that individual micro-regions of a single metasurface chip could be addressed independently, paving the way for spatially multiplexed, multi-wavelength coherent arrays integrated on a single monolithic substrate.

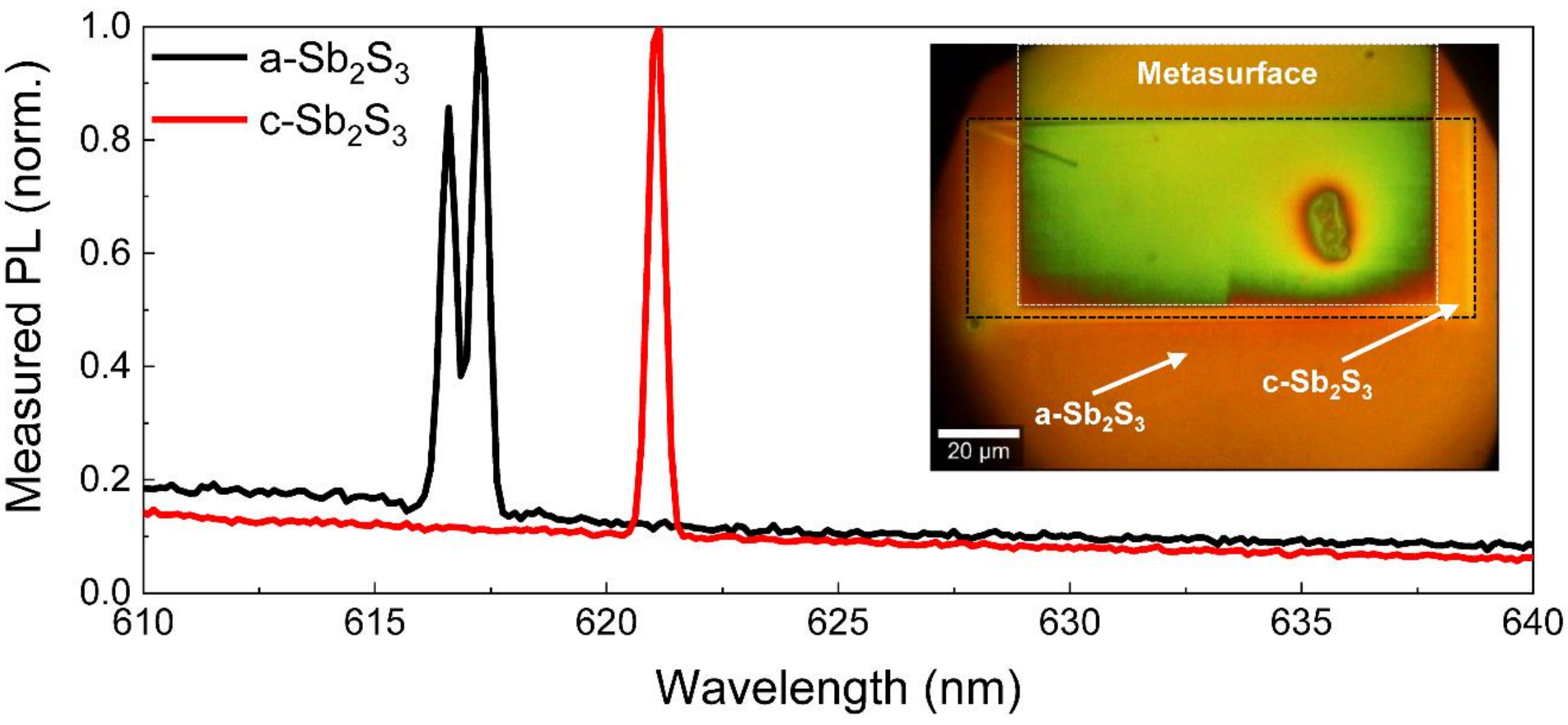

**Figure 4.** Spectral tuning of the emission from the all-optically controlled hybrid BIC nanolaser. The main panel displays normalized visible PL spectra for both the amorphous (a-$Sb_2S_3$) and crystalline (c-$Sb_2S_3$) phases, demonstrating a clear lasing wavelength shift from approximately 616 nm to 621 nm under ns-laser excitation above the lasing threshold. (Inset) WiTec optical microscopy image capturing the phase transition region. The black dashed outline indicates the localized c-$Sb_2S_3$ area induced by continuous-wave (CW) laser exposure, contrasting with the surrounding unswitched a-$Sb_2S_3$ background. The white dashed line marks the metasurface boundary, highlighting the high-resolution, high-speed spatial precision of the all-optical tuning mechanism.

The true versatility of the platform is fully realized during the re-amorphization process. Transitioning back to the amorphous state requires a fast "melt-quench" process, wherein the material must be rapidly heated above ($T_m$) and allowed to cool at a rate faster than the critical crystallization velocity. To achieve this without ablating the nanostructure, a picosecond (ps) pulsed laser operating at 1060 nm in the NIR is utilized. The selection of these specific tuning wavelengths for CW at 405 nm and ps-pulsed at 1060 nm is essential for thermal management of tunable BIC cavity. Both tuning wavelengths are explicitly engineered to lie completely outside the resonant absorption band of the Rd B gain medium, which is exclusively pumped by the 532 nm source. By decoupling the optical channels for structural tuning from the channels for optical pumping, the organic dye is entirely protected from catastrophic multiphoton absorption and localized photo-thermal bleaching during the high-energy melt-quench cycle. This optical transparency, coupled with the rigorous Kapitza thermal resistance provided by the underlying $Al_2O_3$ spacer, ensures the long-term stability and cyclic endurance of the hybrid nanolaser as illustrated in Figure S3.

Moving beyond the generation of tunable coherent light, the metasurface architecture provides the unique capability to simultaneously manipulate the spatial wavefront of the emitted beam, effectively merging a laser cavity and a bulk optical lens into an ultra-thin, singular planar device. To realize this nonlocal light-emitting metalens (Figure 5a), the strictly periodic arrangement of the meta-atoms was replaced with a spatially varying phase gradient (Figure S6). The implementation of metalens phase mask within an active cavity is illustrated in Section 6 in SI. The geometric parameters of each meta-atom must be meticulously tuned to multiplex two entirely distinct optical phenomena: they must provide the correct local phase retardation required by the metalens equation to achieve a high numerical aperture (NA = 0.9), while simultaneously satisfying the strict multipolar interference conditions necessary to maintain the high-$Q$ quasi-BIC resonance for lasing action across the entire array. Figure 5b displays the calculated two-dimensional spatial phase mask required for this dual functionality. Electrodynamic simulations of the resulting metalens-based nanolaser evaluate its performance at the focal plane. In the initial a-$Sb_2S_3$ state (Figure 5c), the 2D electric field intensity profiles ($|E_{x\text{-}y}|^2$ and $|E_{x\text{-}z}|^2$) reveal a tightly concentrated focal spot. The generated coherent light at $\lambda_{a\text{-}Sb_2S_3}$ is not emitted as a divergent profile but rather is instantly collimated and focused by the inherent structural gradient of the cavity itself, demonstrating near-diffraction-limited focusing capabilities. The robustness of this simultaneous lasing and wavefront shaping is further validated upon undergoing the structural phase transition. As demonstrated in Figure 5d, when the underlying $Sb_2S_3$ layer is crystallized, shifting the coherent emission to $\lambda_{c\text{-}Sb_2S_3}$, the macroscopic performance of the metalens is preserved. The 2D focal plane intensity maps reveal that the focal length *f* remains incredibly stable despite the spectral red-shift of the emission. The device successfully maintains its high-NA, diffraction-limited focusing in both the *x-z* and *x-y* planes. This remarkable stability indicates that the active perturbation introduced by the PCM alters the local phase-matching condition for the quasi-BIC just enough to shift the lasing frequency, but the macroscopic spatial phase gradient dictating the wavefront remains dominant and functionally intact. This results in a truly versatile, programmable nanophotonic component: a monolithic, flat-optic laser that can be non-volatilely toggled between emission wavelengths while actively projecting its output onto a deterministic spatial coordinate without the requirement for bulky external optics.

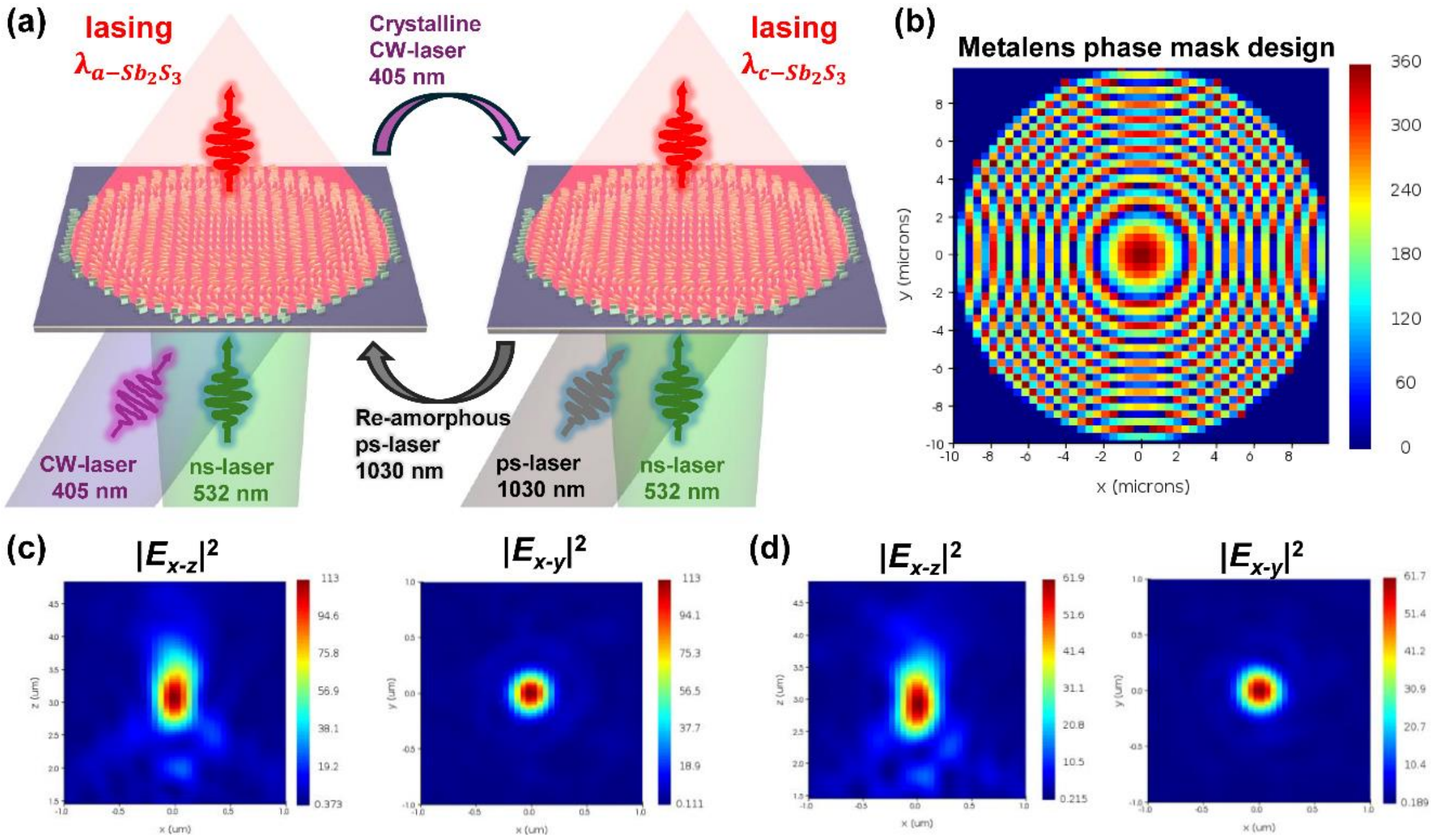

**Figure 5.** Design and validation of the all-optically reconfigurable, BIC-driven metalens nanolaser. (a) (Left) Three-dimensional schematic of the hybrid device utilizing a 405 nm CW pump laser to induce PCM crystallization, alongside a 532 nm ns-laser to pump the gain medium, yielding tunable visible lasing in transmission. (Right) Corresponding 3D schematic illustrating the re-amorphization of the $Sb_2S_3$ layer using a 1060 nm ps-laser, while maintaining the 532 nm ns-laser pump for emission. (b) Spatial phase profile of the metalens engineered to provide strong focusing with a numerical aperture (NA) of 0.9. (c) Simulated optical performance of the metalens nanolaser in the amorphous phase. Calculations executed at $\lambda_{a\text{-}Sb_2S_3}$ display the 2D electric field intensities across both the *x-z* plane ($|E_{x\text{-}z}|^2$) and the *x-y* plane ($|E_{x\text{-}y}|^2$) at the focal point, confirming diffraction-limited focusing capabilities. (d) Simulated optical performance following transition to the crystalline state. Evaluated at $\lambda_{c\text{-}Sb_2S_3}$, the focal plane intensity distributions demonstrate that the focal length remains stable during the emission shift from $\lambda_{a\text{-}Sb_2S_3}$ to $\lambda_{c\text{-}Sb_2S_3}$, maintaining a tight, diffraction-limited focal spot in both the *x-z* and *x-y* cross-sections.

## CONCLUSION

The study presents an all-optically reconfigurable nanolaser operating at room temperature within the visible spectrum. The device combines a high-$Q$ BIC resonance in a $Nb_2O_5$ metasurface with a Rhodamine B dye layer to achieve lasing. By integrating a low-loss $Sb_2S_3$ phase-change material, the lasing wavelength is actively and non-volatilely tuned from 616 nm to 621 nm. A dual-pump scheme (532 nm for lasing, 1060 nm for tuning) and an $Al_2O_3$ thermal barrier ensure stable operation without photothermal damage. The approach also introduces a nonlocal light-emitting metalens that unifies lasing and wavefront control in a single planar device. Beyond tunable lasing, this architecture serves as a prototype for all-optical neurons in neuromorphic photonic computing.[50] The lasing threshold acts as a nonlinear activation function, while the phase-change state stores optical synaptic weights, enabling all-optical matrix-vector multiplication. Additionally, replacing the dye with quantum emitters could yield a reconfigurable quantum light source with Purcell-enhanced emission and programmable photon steering, offering a promising route toward integrated, artificially intelligent, and ultrafast nanophotonic and quantum photonic platforms.[51-57]

## METHODS

### Transmission and Field Profile Simulations

Numerical simulations of the optical properties were performed utilizing a 3D finite-difference time-domain (FDTD) commercial solver (Lumerical FDTD Solutions).[58] The complex refractive indices for all materials such Rd B dye, $Nb_2O_5$, $Sb_2S_3$, $Al_2O_3$, and $SiO_2$ (see Figure S1, SI) were based on experimental data obtained from ellipsometry measurements of the deposited films. The hybrid metasurface cavity was excited by a plane wave source, linearly polarized along the *x*-direction. To simulate an infinite array, periodic boundary conditions (PBC) were applied to the horizontal (*x*-*y*) boundaries, while perfectly matched layers (PML) terminated the vertical (*z*-axis) simulation volume to absorb scattered light. A high-resolution mesh of below 5 nm or finer was implemented to accurately resolve the nanostructured features. Transmission spectra were captured by a field monitor placed in the quartz substrate, and field profiles at the BIC resonances were recorded by 2D monitors bisecting the $Nb_2O_5$ and $Sb_2S_3$ polygon pillars. A stringent auto-shutoff criterion of $1\times10^{-7}$ was used to ensure simulation convergence to a high-accuracy steady state.

**Multipolar decomposition calculations**

To identify the nature of the resonant modes, multipolar decomposition was performed within the Lumerical FDTD environment. This calculation required two 3D field monitors placed around the metasurface cavity meta-atom. The first monitor was used to compute the electric and magnetic field components at each mesh point within the nanostructures, while the second was used to determine the effective refractive index of the cavity volume. The scattering cross-sections for the electric dipole ($C_{ED}$), electric quadrupole ($C_{EQ}$), magnetic dipole ($C_{MD}$), and magnetic quadrupole ($C_{MQ}$) were then calculated using equations below. The detailed expression for each physical parameter in equations for these scattering power components are available in the literature.[59]

$$C_{ED} = \frac{k_0^4}{6\,\pi\epsilon_0^2 E_0^2}\left|p_{car} + \frac{ik_0}{c}\left(t + \frac{k_0^2}{10}\overline{R_t^2}\right)\right|^2 \quad (1),$$

$$C_{EQ} = \frac{k_0^6}{80\,\pi\epsilon_0^2 E_0^2}\left|\overline{\overline{Q_e}} + \frac{ik_0}{c}\overline{\overline{Q_t}}\right|^2 \quad (2),$$

$$C_{MD} = \frac{\eta_0^2 k_0^4}{6\,\pi E_0^2}\left|m_{car} - k_0^2\overline{R_m^2}\right|^2 \quad (3),$$

$$C_{MQ} = \frac{\eta_0^2 k_0^6}{80\,\pi E_0^2}\left|\overline{\overline{Q_m}}\right|^2 \quad (4),$$

**Fabrication of all optical BICs nanolaser metasurface**

The fabrication process for the tunable hybrid metasurface is outlined in Figure S2. The procedure began with double-sided polished deep-UV quartz substrates (Photonik Singapore), which were cleaned by sequential sonication in acetone and isopropanol (IPA) for 10 minutes, followed by a final rinse and nitrogen-gas drying. A 130 nm thick film of $Sb_2S_3$ was then deposited using RF sputtering (UBM) at room temperature, utilizing Argon plasma (21.6 sccm flow, 20 W RF power, 10 mTorr). A 10 nm $Al_2O_3$ thermal insulation layer was subsequently deposited via atomic layer deposition (ALD, Beneq TFS 200) at 80 °C (4 mTorr chamber pressure), using trimethylaluminium (TMA) and $H_2O$ as precursors over 100 cycles with $N_2$ purges. The Niobium Pentoxide ($Nb_2O_5$) pillar molds were defined using electron beam lithography (EBL, Elionix ELS-7000). A positive-tone resist (ZEP520A) was spin-coated at 3000 rpm for 90 s to achieve a 400 nm thickness, followed by a 180°C bake for 2 minutes. A conductive layer (Ezspacer) was applied (1500 rpm, 30 s) to prevent charging during exposure. The 300×300 $\mu m^2$ patterns were exposed at 100 kV acceleration with a 200 pA beam current and a 320 $\mu C/cm^2$ base dose. After exposure, the Ezspacer was removed with deionized water, and the sample was developed in xylene (30 s) before being quenched in IPA. Next, a 100 nm layer of $Nb_2O_5$ was deposited using ALD (Beneq TFS 200) at 85°C, a temperature below the ZEP520A glass transition. This deposition used a tert-butylimino tris-diethylamino-niobium precursor and $H_2O$, with a growth rate of ~0.59 Å/cycle. A blanket etch was performed using inductively coupled plasma reactive ion etching (ICP-RIE, Oxford OIPT Plasmalab) to remove the top 100 nm of $Nb_2O_5$. The etch utilized $CHF_3$ plasma (25 sccm, 150 W RF, 900 W ICP, 25 mTorr) at a rate of 150 nm/min. The remaining ZEP520A resist was stripped using an $O_2$ plasma descum (50 sccm $O_2$, 250 W RF, 3 min). Finally, to enhance gain medium adhesion, the sample was exposed to UV light in a vacuum for 10 minutes. Immediately following this, the gain medium, consisting of Rhodamine B dye dissolved in polyvinyl alcohol (PVA), was spin-coated onto the metasurface.

**Transmission measurement**

Far-field transmission spectra were acquired using a supercontinuum nanosecond laser (Leukos Laser Inc., Opera) with a 30 kHz repetition rate. The incident light was conditioned to be *x*-polarized, matching the resonance, by passing it through a sequence of a quarter-wave plate (QWP), a half-wave plate (HWP), and a linear polarizer (LP). The light transmitted through the sample was collected by an optical fiber and analyzed with an Ocean Optics USB4000 spectrometer.

**Lasing measurement**

Lasing characterization was performed using the photoluminescence (PL) setup shown in **Figure S6**. The pump source was a free-space nanosecond (ns) laser operating at 532 nm (10 Hz repetition rate, avg. power up to 50 mW, pulse energy up to 150 μJ). The incident pump power was controlled with a variable attenuator. A linear polarizer (LP) and half-wave plate (HWP) were used to set the linear polarization of the pump to match the metasurface resonance. The pump beam was focused onto the sample (spot diameter > 50 μm) using a focusing lens (FL). The sample was mounted on a motorized *XYZ*-stage for 2D scanning and *z*-axis focusing. Emission was collected via a collecting lens (CL), and a 532 nm notch filter was used to block the pump laser from the collected signal. A white light source, beam splitter (glass sheet), and CCD camera (with flipping mirror M1 and lens L1) were used for sample alignment and imaging. The lasing spectra were captured by a WiTEC PL detector equipped with a 150 or 600 grooves/mm grating.

## ASSOCIATED CONTENT

### Supporting Information

The measured ellipsometer refractive indices of Rd B, $Nb_2O_5$, $Sb_2S_3$, and $Al_2O_3$; nanofabrication flow for the tunable nanolaser metasurface cavity; PL of Rd B dye; Lasing measurement setup; Benchmark comparison table with literature works for nanolaser, phase gradient light emitting metasurface, and tunability using nanophotonic cavities.

## CORRESPONDING AUTHORS

Email: Omar_Abdelrahman@a-star.edu.sg

## ORCID

Omar A. M. Abdelraouf: https://orcid.org/0000-0002-9065-7414

## AUTHOR CONTRIBUTIONS

O.A.M.A. conceived the physical concept, idea of tunable lasing, designed the active all optical BIC nanolaser cavity, carried out 3D FDTD simulations and multipolar decomposition, did materials deposition and optical characterization, cleanroom nanofabrication of the all optical BIC nanolaser cavity, linear optical measurements, photoluminescence and lasing measurements, optical reconfigurable of phase change materials, and wrote the whole manuscript.

### Competing Financial Interests

The authors declare no competing financial interest.

**ACKNOWLEDGMENT**

Authors acknowledge funding from A*STAR under its Career Development Fund (CDF) grant no. C233312016 and SERC Central Research Funds (CRF). Also, the support from Singapore international graduate award (SINGA).